\documentclass[12pt]{iopart}

\usepackage{iopams} 
\usepackage{bm,graphics,graphicx,epsfig,color}

\begin{document}

\title[Bounding the mass of the graviton]{Bounding the mass of the graviton with gravitational waves: Effect of higher harmonics in gravitational waveform templates}

\author{K. G. Arun$^{1,2,3}$ and Clifford M. Will$^{1,3}$}

\ead{arun@physics.wustl.edu, cmw@wuphys.wustl.edu}
\address{
$^1$
GReCO, Institut d'Astrophysique de Paris, UMR 7095-CNRS,
Universit\'e Pierre et Marie Curie, 98$^{bis}$ Bd. Arago, 75014 Paris, France \\
$^2$ 
 LAL, Universit\'e Paris Sud, IN2P3/CNRS, Orsay, France \\
$^3$ McDonnell Center for the Space Sciences, Department of Physics,
Washington University, St. Louis MO 63130 USA}

\begin{abstract}
Observations by laser interferometric detectors of gravitational waves from inspiraling compact binary systems can be used to search for a dependence of the waves' propagation speed on wavelength, and thereby to bound the mass or Compton wavelength of a putative graviton. We study the effect of including higher harmonics, as well as their post-Newtonian amplitude corrections, in the template gravitational waveforms employed in the process of parameter estimation using matched filtering. We consider the bounds that could be achieved using advanced LIGO, a proposed third generation instrument called Einstein Telescope, and the proposed space interferometer LISA.  We find that in all cases, the bounds on the graviton Compton wavelength are improved
by almost an order of magnitude for higher masses when amplitude corrections
are included.
\end{abstract}
\pacs{04.30.-w, 04.80.Cc and 04.80.Nn}

\maketitle

\section{Introduction and summary}

The ongoing development of laser interferometric gravitational-wave observatories on the ground and in space brings closer the day when gravitational radiation will be used as a tool for astronomical discovery and for testing fundamental physics~\cite{SathyaSchutzLivRev09}.  
The Laser Interferometer Gravitational-Wave Observatory (LIGO)~\cite{ligo} recently finished its fifth science run, operating for a year and half at its initial design sensitivity. 
Its European counterpart Virgo~\cite{virgo} ran for four months in coincidence with 
the final period of LIGO's science run.  Both detectors will undergo a series of upgrades to improve their sensitivity by a factor of about ten and to be back on the air in the 2014 time frame.  A third generation European interferometer provisionally called Einstein Telescope (ET)~\cite{ET} is being planned to have unprecedented low-frequency sensitivity, with a seismic cutoff close to 1 Hz.  While the ground-based
detectors will be sensitive to high-frequency gravitational waves ($ 1-10^4$ Hz), the proposed Laser Interferometer Space Antenna (LISA)~\cite{Danzmann97}, will be sensitive to low-frequency gravitational waves in the milli-Hertz range. LISA will therefore be able
to detect gravitational waves from sources like supermassive black hole (SMBH) binaries and will complement the ground-based detectors which are to be sensitive to stellar mass and intermediate mass black hole (BH) binaries. Since the gravitational waves from these compact binaries
can be very precisely modelled using analytical and numerical relativity, the detection and parameter estimation will be performed by the technique of matched filtering, whereby theoretically generated waveforms are used as templates to search for
gravitational-wave signals in the data.

In previous papers~\cite{Will98,WillYunes04,BBW05a,BBW05b} we studied
the extent to which observations of such gravitational waves could test gravitational theory, in particular test whether the speed of gravitational waves depends on their wavelength, or, to use a shorthand phrase, whether the graviton has a mass (the waves themselves are completely describable by non-quantum physics, of course).  We found that the upgraded ground-based detectors, provisionally denoted by AdvLIGO and AdvVirgo, could place a lower bound on the graviton Compton wavelength comparable to bounds ($\sim 10^{12} \, {\rm km}$) obtained from solar system dynamics, while LISA could do some four orders of magnitude better.   

The basic idea is simple:\ if there is a mass associated with the propagation of gravitational waves (``a massive graviton''), then the speed of propagation will depend on wavelength in the form
$v_g \approx 1 - (\lambda/\lambda_g)^2$, where $\lambda_g$ is the Compton wavelength of the graviton, in the limit where $\lambda \ll \lambda_g$.  Irrespective of the nature of the alternative theory that predicts a massive graviton (and notwithstanding the difficulties in defining such theories free of pathologies such as the ZvDV discontinuity~\cite{Zakharov70,Veltman70}), it is reasonable to expect the differences between such a hypothetical theory and general relativity in the predictions for the evolution of massive compact binaries to be of order $(\lambda/\lambda_g)^2$, and therefore to be very small, given that $\lambda \sim 10^3$ km for stellar mass inspirals and $\sim 10^8$ km for massive black hole inspirals.

As a result, the gravitational waveform seen by an observer close to the source will be very close to that predicted by general relativity.  However, as seen by a detector at a distance $D$, hundreds to thousands of Mpc away, the phasing of the signal will be distorted because of the shifted times of arrival, $\Delta t \sim D(\lambda/\lambda_g)^2$ of  waves emitted with different wavelengths during the inspiral.   In addition to measuring the astrophysical parameters of the system, such as masses and spins,
the matched filtering technique permits one to estimate or bound such effects.
We point out that small deviations from general relativity will not strongly
affect the detection of the gravitational wave signals. The reason is that,
for detection, one maximizes the overlap function of the signal with
templates over the parameters of the template (see \cite{DIS98} for details). Even
if the actual signal is slightly different from that predicted by GR,
templates based on GR should be able to recover it but with significant
biases in the estimated parameters. Parameter estimation hence should be
seen as a post-detection problem.

In our earlier  matched filtering analyses~\cite{Will98,WillYunes04,BBW05a,BBW05b}, we chose a particular form for the theoretical waveform template, known as the ``restricted waveform (RWF)'', constructed using the dominant quadrupole amplitude of the wave evaluated in the lowest-order ``Newtonian'' approximation, but with a phase expressed to the highest post-Newtonian (PN) order available at the time.  This was generally second post-Newtonian (2PN) order, or $O([v/c]^4)$ beyond the leading quadruple approximation,  and included the effects of non-precessing spin (see \cite{Bliving} for a review of the post-Newtonian phasing formulae).  The Fourier domain
waveform in this approximation has only the leading order amplitude multiplying a term whose phase is proportional
to twice the orbital phase.

However, recent work has pointed out that the inclusion of other multipoles in the wave amplitude, as well as PN corrections to the amplitudes, can have a dramatic effect on the estimation of certain system parameters~\cite{SinVecc00a,SinVecc00b,MH02,HM03,Chris06,ChrisAnand06,ChrisAnand06b,AISS07,AISSV07,TriasSintes07,PorterCornish08,AMVISS08}.
The effects of incorporating higher harmonics include increased mass reach for the ground-based~\cite{ChrisAnand06} and space-based~\cite{AISS07} detectors,
improved estimation of the mass parameters~\cite{ChrisAnand06b,AISSV07,TriasSintes07,PorterCornish08}, improved angular resolution~\cite{MH02,AISSV07,TriasSintes07,PorterCornish08} and improved estimation of cosmological parameters such
as those associated with the dark energy equation of state~\cite{AISSV07,AMVISS08}.
 
These improvements arise from two effects.  (i) The presence of higher harmonics of the orbital phase increases the
frequency span of the signal in the detector band. (ii) The structure of
the waveform is richer because, even though they are smaller than the dominant quadrupole term, the new amplitude terms and their PN corrections introduce new functions
of masses, spins, and inclination angles.  

The purpose of this paper is to explore the effects of using the full waveform (FWF) on the bounds on $\lambda_g$. 
In particular, we revisit the earlier bounds for non-spinning binaries by including the effects of
PN amplitudes up to 3PN order~\cite{BIWW96,ABIQ04,KBI07,BFIS08} and phasing up to 3.5 PN order~\cite{BDIWW95,B96,BFIJ02,BDEI04}, including harmonics as high as 8 times the
orbital phase.
We also consider the three different detector
configurations mentioned above: the second generation AdvLIGO, the proposed third generation ET, and the proposed space-based interferometer LISA. 

\begin{center}
\begin{figure}[t]
\includegraphics[scale=0.5]{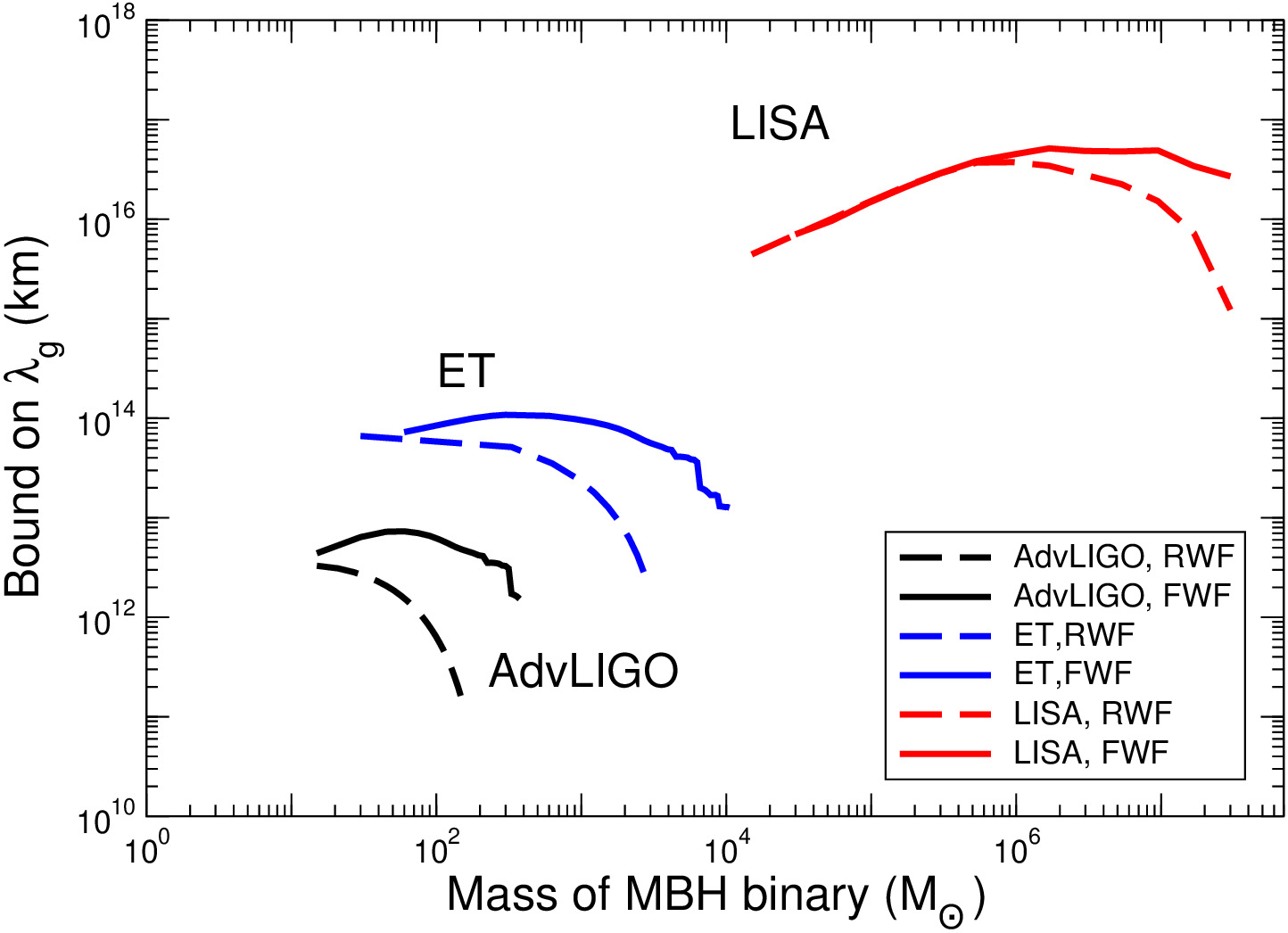}
\label{summary}
\caption{Bounds on the graviton Compton wavelength that can be deduced from AdvLIGO, Einstein Telescope and LISA.  The mass ratio is 2.   The distance to the source is assumed to be
100 Mpc for AdvLIGO and ET, and 3 Gpc for LISA. 
}
\end{figure}
\end{center}

Fig.~\ref{summary} displays our central results. The 1-$\sigma$ bounds on $\lambda_g$ obtainable from AdvLIGO, ET and LISA are plotted as a function of the total
mass of the binary for a fixed mass ratio of $m_2/m_1=2$. For AdvLIGO and ET, the source is assumed to be at a luminosity distance of 100 Mpc and for LISA the SMBH binary is assumed to be 3 Gpc away.   The bounds from the Newtonian RWF and 3PN FWF are compared. Inclusion of amplitude corrections and the higher harmonics improve the bounds for both ground-based
configurations and at the high-mass end for LISA. The improvement is more than an order of magnitude for heavier
binaries, because higher harmonics play a more prominent role for such systems.
Typical bounds, with the use of higher harmonics, for AdvLIGO, ET and LISA are $10^{12}$ km, $ 10^{13}$ km and $10^{16}$ km, respectively.  The best bound, not surprisingly, will be provided by LISA, thanks to its low frequency sensitivity, to the high signal-to-noise ratios with which it will be observing the supermassive binary black hole
coalescences, and to the very large distances involved. 
Though our results are for a specific location and orientation of the binary, we have verified that the bounds are not significantly altered by different source positions and orientations.

The remainder of the paper provides details underlying these results.  In Sec. \ref{sec:assumptions}, we describe the full-waveform model used, the noise curves for the various detectors, and the technique of matched filtering.  Section \ref{sec:bounds} details the bounds obtainable from the various detectors.

\section{Parameter estimation using full waveform templates}
\label{sec:assumptions}

As our waveform model we begin with amplitude-corrected, general relativistic waveforms which are 3PN accurate in amplitude
and 3.5PN accurate in phasing.  We ignore the spins of the bodies in the binary system.
Previous calculations used waveforms which are of Newtonian order in amplitude
and 2PN order in phase.  As opposed to the Newtonian waveforms, the 3PN amplitude-corrected waveforms contain all harmonics from $\Psi$ up to $8 \,\Psi$, where $\Psi$ is the
orbital phase (the leading quadrupole component is at $2\Psi$). 

The effect of a massive graviton is included in the expression
for the orbital phase following Ref.~\cite{Will98}.  The wavelength-dependent propagation speed changes the arrival time $t_a$ of a wave of a given emitted frequency $f_e$ relative to that for a signal that propagates at the speed of light; that time is given, modulo constants,by
\begin{equation}
t_a = (1+Z) \left [ t_e + \frac{D}{2\lambda_g^2 f_e^2} \right ] \,,
\label{time}
\end{equation}
where $f_e$ and $t_e$  are the wave frequency and time of emission  as measured at the emitter, respectively, $Z$ is the cosmological redshift, and
\begin{equation}
D \equiv {(1+Z) \over a_0}\int_{t_e}^{t_a} a(t) dt \,,
\label{D}
\end{equation}
where $a_0=a(t_a)$ is the present value of the scale factor (note that $D$ is not exactly the luminosity distance~{\footnote{For $Z\ll1$, $D$ is roughly equal to luminosity distance $D_L$. Hence we have assumed $D\simeq D_L$  in the
case of ground based detectors for which we consider sources at 100 Mpc. For LISA, we have carefully accounted for this difference. }}).  This affects the phase of the wave accordingly.  In the frequency domain, this adds a term to the phase $\psi(f)$ of the Fourier transform of the waveform given by $\Delta \psi(f) = -\pi D/f_e \lambda_g^2$.   Then, for each harmonic of the waveform with index $k$, one adds the term 
\begin{equation}
\Delta \psi_k (f) = \frac{k}{2} \Delta \psi(2f/k) = - \frac{k^2}{4} \pi  D/f_e \lambda_g^2 \,.
\label{phaseterm}
\end{equation}
Here $k=2$ denotes the dominant quadrupole term, with phase $2\Psi$, $k=1$ denotes the term with phase $\Psi$, $k=3$ denotes the term with phase $3\Psi$, and so on.

This is an adhoc procedure
because a massive graviton theory will undoubtedly deviate from GR not just in the propagation effect, but also in the way gravitational wave damping affects the phase, as well as in
in the amplitudes of the gravitational waveform.  If, for example, such a theory introduces a leading correction to the quadrupole phasing $\psi_{\rm quad} \sim (\pi {\cal M} f_e)^{-5/3}$ of order $(\lambda/\lambda_g)^2 \times (\pi {\cal M} f_e)^{-5/3}$, where $\cal M$ is the chirp mass, then the propagation induced phasing term (\ref{phaseterm}) will be larger than this correction term by a factor of order $k^2 (D/{\cal M})(\pi {\cal M} f_e)^{8/3} \sim (D/{\cal M})v^8$.  Since $v \sim 0.1$ for the important part of the binary inspiral, and $D \sim$ hundreds to thousands of Mpc, it is clear that the propagation term will dominate.  In any case, given the fact that
there is no generic theory of a massive graviton, we have no choice but to omit
these unknown contributions.  

However the effect of our assumption of neglecting the spin of the binary could be more severe. As studied in detail in~\cite{BBW05a}, the spin effects could
weaken the bounds on $\lambda_g$. This conclusion was based on the restricted
waveforms and on nonprecessing spins. When the higher harmonics from the
polarization as well as spin precession are included, the trends may be different. But this will require a thorough analysis of parameter estimation with precessing binary waveforms; this will be the subject of future work.  

\subsection{Instrument noise models}

The matched filtering procedure for estimating errors requires knowledge of the noise characteristics of the detectors.   In the frequency domain, these characteristics are embodied in the noise power spectral density (PSD).
We use analytical fits to the designed PSDs
of various detector configurations.  For AdvLIGO,
we have used the analytical fit provided in Eq.~(3.7) of ~\cite{AISS05}.
We assume the ET to be
an L-shaped detector with an arm length of 10 km. 
The PSD we use is from Ref.~\cite{ETPSD}. This is essentially
the analytical fit given in  Eqs (4.4) and (4.5) of Ref.~\cite{ChrisAnand06b}
with slight modifications to incorporate the arm length.

For the case of LISA, we have used the noise PSD from Ref.~\cite{BBW05a},
Eqs (2.28)--(2.32),
though there have been minor modifications to the LISA noise PSD since then.
This choice of the noise curve helps us to calibrate our codes by reproducing the results of Ref.~\cite{BBW05a} for the restricted waveform case.

\subsection{Frequency cut-offs}

The gravitational wave signals are converted from the time domain to the frequency domain using the stationary-phase approximation (SPA).
We assume the signals to last in the respective detector bands between
frequencies $f_{\rm lower}$ and $f_{\rm upper}$.   We assume that the upper cut-off is given by the last stable orbit (LSO) of the system.  Hence the $k-$th harmonic, for instance, is sharply cut-off at $k\,F_{\rm LSO}$ using step functions, where $F_{\rm LSO}$ is the orbital frequency at the last stable orbit.
Readers can refer
to Refs.~\cite{ChrisAnand06,AISS07,ABFO08} 
 for the details and subtleties associated with this method and related assumptions. See Refs~\cite{DPK99,DIS00} for a general discussion of the step function
cut-offs employed in SPA. 

For the ground-based detectors,
the low frequency cut-off
 is fixed by the seismic cut-off of the detector.
Following the designed noise PSDs, we assume
this to be 20 Hz for AdvLIGO and 1 Hz and 10 Hz for Einstein Telescope. 

 For the
case of LISA, the low frequency cut-off is defined in a different way. Since LISA can observe the sources for extended periods of order months to years, we make the standard assumption that the source is observed
in the LISA band for one year prior to merger. For a particular SMBH system, we use $F_{\rm LSO}$ to be a measure of the upper cut-off orbital frequency and deduce the low orbital frequency cut-off assuming a 1 year  (or less) duration of the signal in the LISA band.   Hence both the low frequency and high frequency cut-offs will be different for each harmonic (see Ref.~\cite{AISS07} for details).
Following Ref.~\cite{BBW05a}, we choose
\begin{eqnarray}
f_{\rm lower}&=&{\rm Max} \left(10^{-4}{\rm Hz},\,f_{\rm begin}\right),
{\rm where}\\
f_{\rm begin}&=& 4.149\times10^{-5} \left(\frac{{\cal M}}{10^6 M_\odot}\right)^{-5/8}
\left(\frac{T_{\rm obs}}{1 {\rm yr}}\right)^{-3/8} \,,
\end{eqnarray}
where ${\cal M}$ is the chirp mass of the binary and $T_{\rm obs}$ is the duration
of observation of the signal, which is assumed to be 1 year (or less). Notice also that
we have assumed, rather conservatively, that LISA is not sensitive to frequencies below $10^{-4}$ Hz.   In our choice the $k$-th harmonic  will last from $kf_{\rm lower}/2$ until $k F_{\rm LSO}$.

\subsection{Assumptions about the sources}

We have assumed the typical distance to the binary black holes to be 100 Mpc for the AdvLIGO
and ET cases, and 3 Gpc for LISA. Notice, however, that the bounds
on $\lambda_g$ are roughly independent of the distance of the source for a given detector~\cite{Will98}.
This is because the size of the massive graviton term in the phasing, Eq.\ (\ref{phaseterm}), is directly proportional to the distance of the source, while the errors in estimating or bounding this term are roughly inversely proportional to signal-to-noise ratio, and therefore also increase linearly with distance.   

All the plots which we provide assume a particular source position and orientation of the binary. These fixed source coordinates are specified with respect to an earth-based coordinate system for the ground based detectors, while for LISA they are specified with respect to the fixed LISA frame (we ignore LISA's orbital motion).

The bounds we derive will vary with the source's location and orientation 
for all three detector configurations.  We performed a numerical experiment to estimate crudely this uncertainty in our estimate by deriving estimates for 100 random realizations of the source direction.  We found, for example, that the bound on $\lambda_g$ varied between $1$ and $8\times10^{12}$ km for AdvLIGO
for a $40M_\odot-80M_\odot$ system.
Hence we expect the typical orders of magnitude we quote to be representative of the bounds obtainable on average.
Note that all the bounds that we quote are for single GW events.  In the
happy event of numerous detections, then the bounds could improve either in
value or in confidence level; however we have not analyzed this possibility.

\subsection{Errors obtained using the Fisher matrix approach}

Our estimate of the bounds on the massive graviton parameter is based
on the Fisher matrix formalism.  We construct the Fisher matrix for the different
detector noise PSDs using the amplitude corrected PN waveform model
described earlier, converted to the 
Fourier domain using the stationary phase approximation.
We use a six-dimensional parameter space consisting of
the time and phase ($t_c, \, \phi_c$) of coalescence, the chirp mass ${\cal M}$, the mass asymmetry parameter $\delta=|m_1-m_2|/(m_1+m_2)$, the massive graviton parameter $\beta_g=\pi^2 D {\cal M}/\lambda_g^2(1+Z)$, and the luminosity distance $D_L$.  We fix the three angles, $\theta$, $\phi$ and $\psi$  which appear in the antenna pattern functions to be $\pi/3$, $\pi/6$ and $\pi/4$ respectively and the inclination angle of the binary to be $\iota=\pi/3$.
 Details of the Fisher matrix approach as applied to the compact binary coalescence signals can be found in Refs.~\cite{CF94,PW95,AISS05}, and more recently
in Ref.~\cite{Vallisneri07} ,which critically reexamines the caveats involved in
using the Fisher matrix formalism to deduce error bounds for various gravitational wave detector configurations. 

The square root of each of the diagonal entries in the inverse of the Fisher matrix gives a lower bound
on the error covariance of any unbiased estimator.  Our focus here is solely
on the diagonal element corresponding to the massive graviton parameter.  The
$1-\sigma$ error bar on $\beta_g$ can be translated into a bound on the 
Compton wavelength using $\Delta \beta_g=\beta_g$, and this is the quantity
that we use in the plots as well as in the discussions.

In the next section we discuss our results in detail for the three detector
configurations considered.

\section{Bounds on massive graviton theories from AdvLIGO, ET and LISA}
\label{sec:bounds}

\subsection{Advanced LIGO and Einstein Telescope}

\begin{center}
\begin{figure}[t]
\includegraphics[scale=0.30] {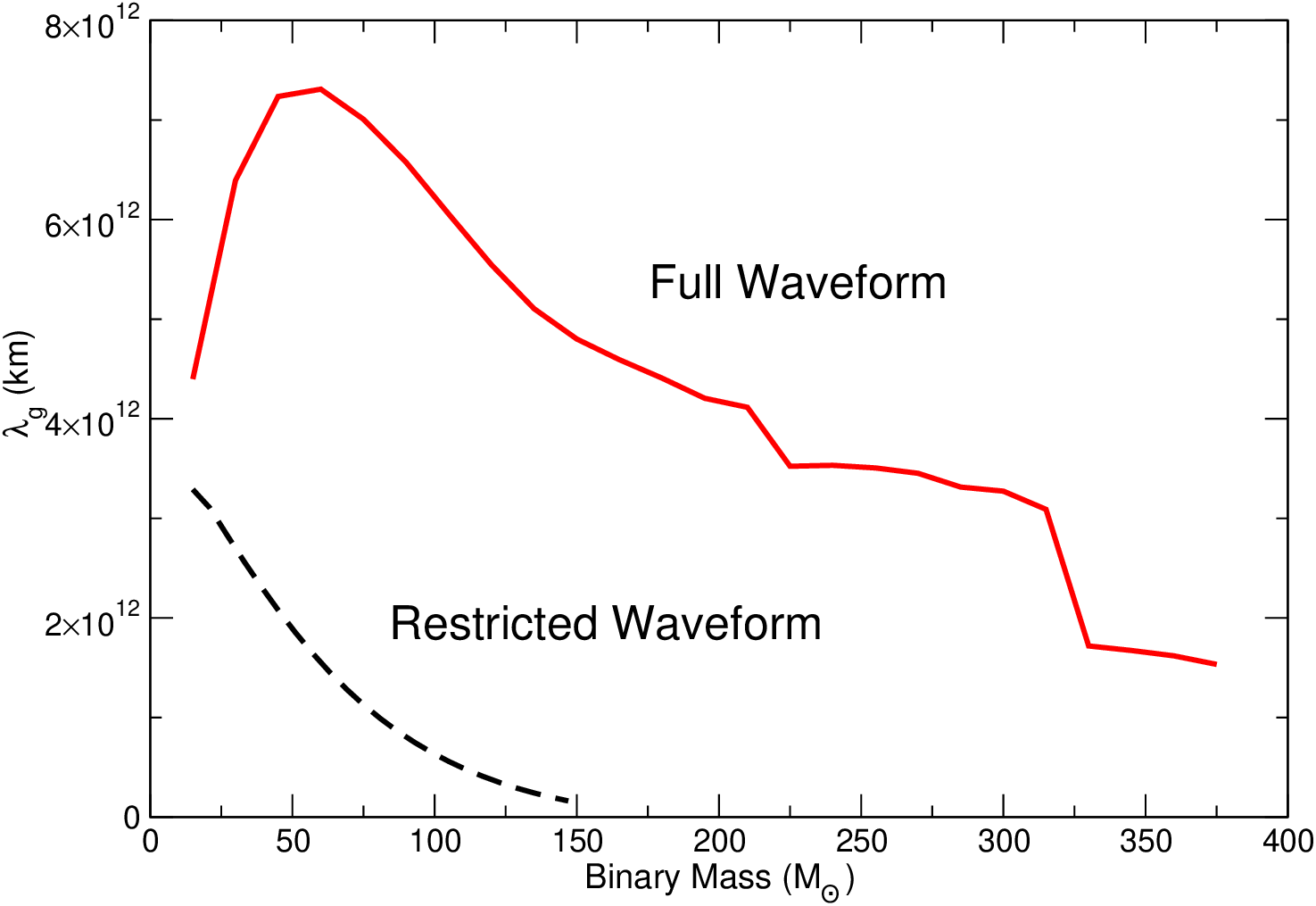}
\hskip 0.2cm
\includegraphics[scale=0.32] {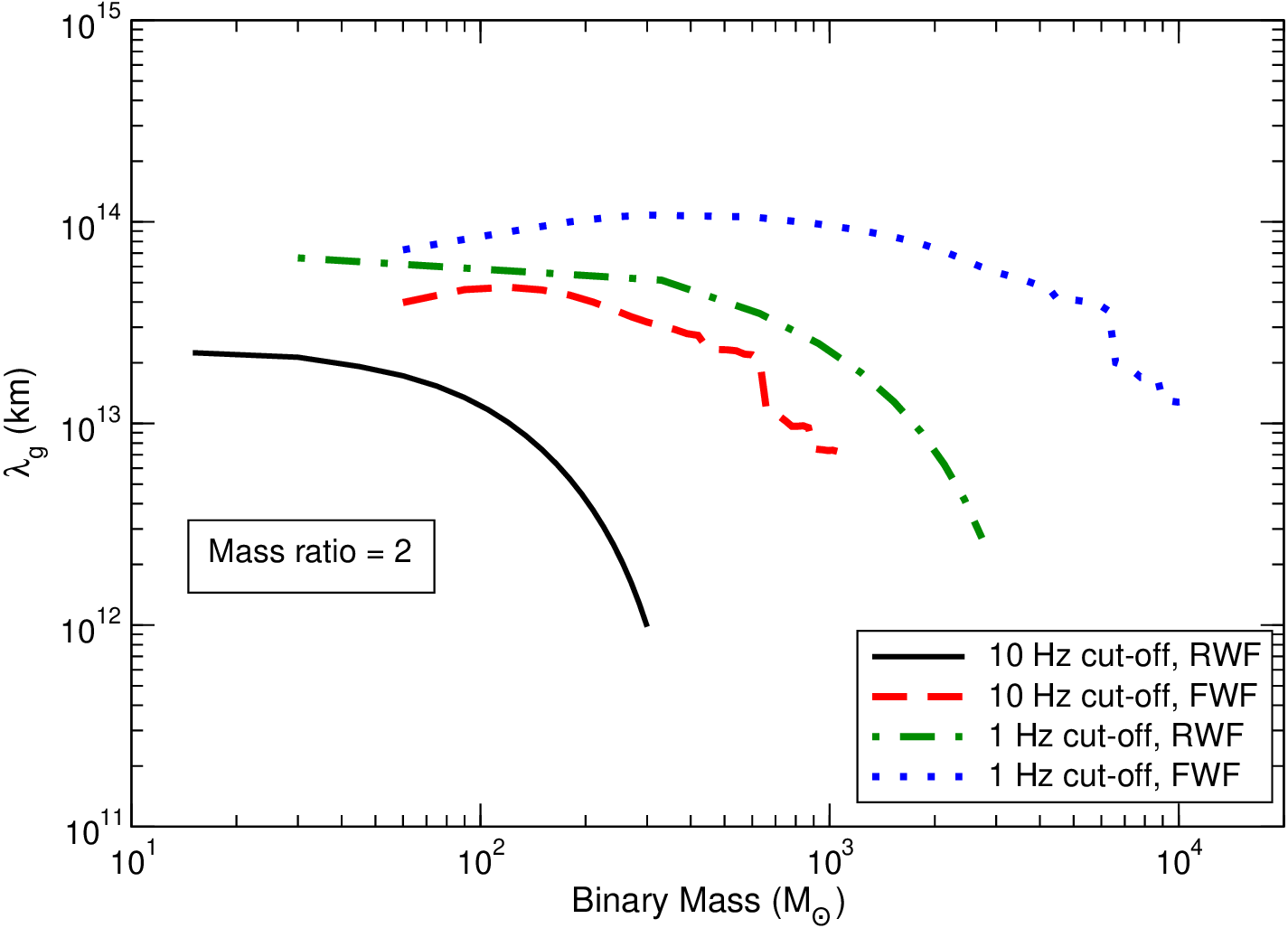}
\caption{Bounds on $\lambda_g$ using restricted and 3PN full waveforms for AdvLIGO (left panel) and Einstein Telescope (right panel). For Einstein Telescope, we show the bounds assuming two possible seismic cut-off frequencies: 10 Hz and 1 Hz. }
\label{fig:AdvLIGO-ET}
\end{figure}
\end{center}

Figure~\ref{fig:AdvLIGO-ET} shows the bounds on $\lambda_g$ possible from future
GW observations with AdvLIGO and Einstein Telescope. 
There is evident improvement from using the FWF in the AdvLIGO case. For binaries
which are well-detected with the RWF, the inclusion of higher harmonics
improves the massive graviton bounds by a factor of a few. The most striking feature
is that even for masses that are beyond the range detectable by RWF templates,
the FWF based templates still put bounds which are better than the best
bounds from RWF. The best bounds would be for intermediate mass BH
binaries whose total mass lies in the range 50-100 $M_\odot$.

Because of its improved
sensitivity, ET will be able to put more stringent bounds, roughly 
an order of magnitude better, compared to AdvLIGO. On the right panel of Fig.~\ref{fig:AdvLIGO-ET}, we show the bounds
that are possible with two different seismic cut-off frequencies, 10 Hz and 1 Hz,
for ET. The FWF results for a 10 Hz cut-off are as good as  the  results for RWF with a 1 Hz cut-off.   This is because including the higher
harmonics produces roughly the same effect as having a 
lower cut-off frequency  because both extend the sensitivity of the detector to lower frequencies. However the best estimates
for ET are still from the case where the seismic cut-off is 1 Hz and
FWF templates are employed and for intermediate mass BH binaries whose
masses are of the order of a few hundreds of solar masses.   Also it is interesting to note
that templates which account for higher harmonics essentially bridge
the gap in mass coverage of different gravitational wave detectors (see
Fig.\ \ref{summary}).
The ``steps'' in the curves for AdvLIGO and ET at the high
mass ends are due to the effects on different harmonics of the sharp step function cut-offs at high frequencies.

In the FWF cases for both AdvLIGO and ET we have checked, for
all mass ranges shown in the plot, that the Fisher matrix is not
ill-conditioned and the binaries are observed with an SNR of at least 8.
There were cases for which the SNR was less than 5 but the Fisher matrix was
found to be well-conditioned, but we have not used those systems because
for very low SNRs, the Fisher matrix formalism itself breaks down~\cite{Vallisneri07}.

\begin{center}
\begin{figure}[t]
\includegraphics[scale=0.5] {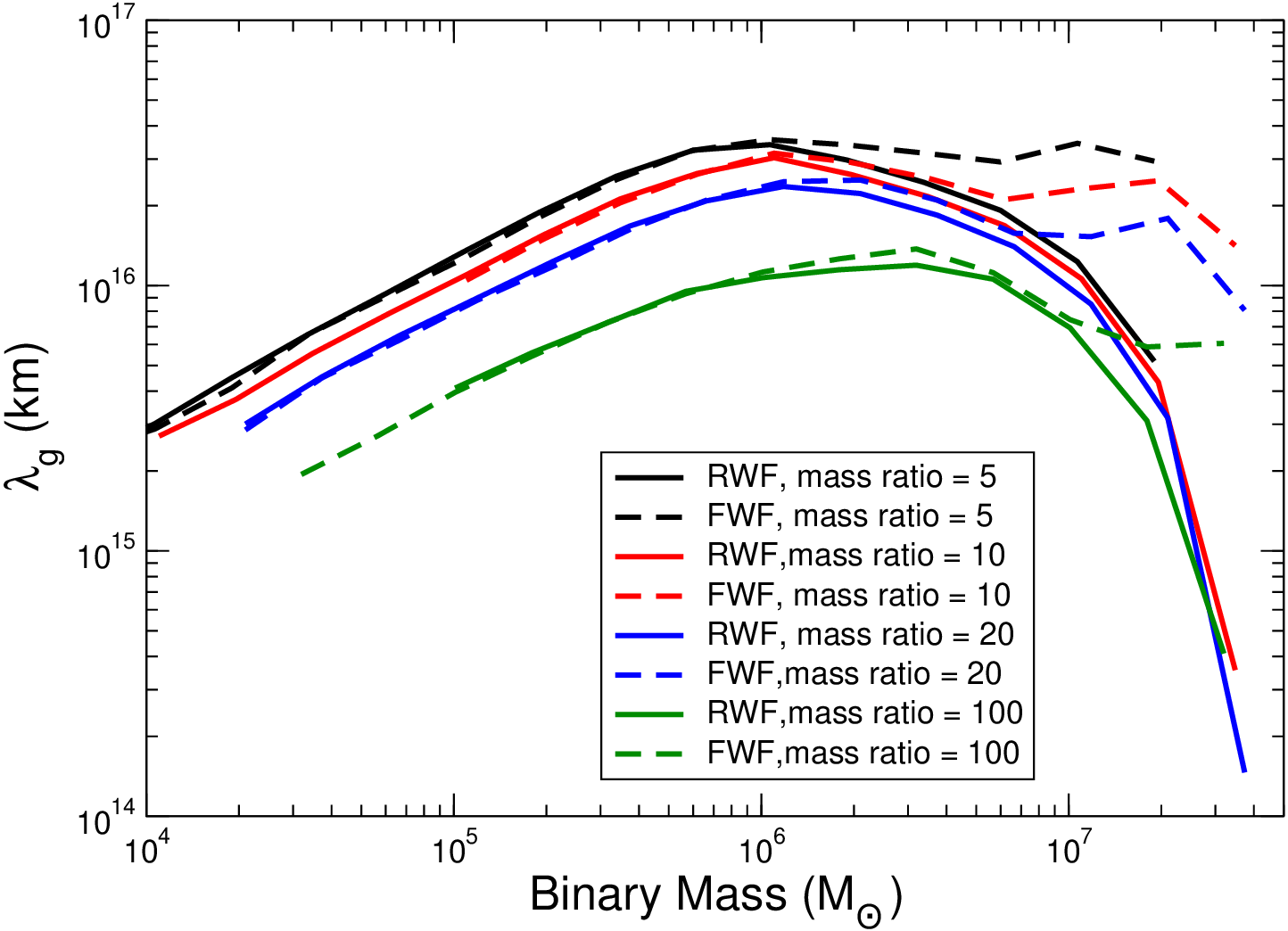}
\caption{Bound on $\lambda_g$ using LISA, as a function of total mass for restricted and full waveforms. Results for different mass ratios 5 --100 are shown. }
\end{figure}
\label{MG-LISA}
\end{center}

\subsection{LISA}

In Fig.\ 3 we show the possible bounds on $\lambda_g$  
from observations by LISA of supermassive binary inspirals for different
mass ratios. The bounds are almost three
orders of magnitude better than those from the best ground-based experiments.
This improvement is almost entirely due to the larger masses involved, as can be seen from the dependence of the bound on $\lambda_g$ on the relevant parameters of the system and the detectors, given by Eq. (4.9) of \cite{Will98}:
\begin{equation}
\lambda_g \propto \left ( \frac{I(7)}{\Delta} \right )^{1/4}
\left ( \frac{D}{(1+Z)D_L} \right )^{1/2} \frac{{\cal M}^{11/12}}{S_0^{1/4} f_0^{1/3}} \,,
\label{lambdappnl}
\end{equation}
where $S_0$ is a parameter that establishes the scale of the PSD (in Hz$^{-1}$), $f_0$ is a characteristic ``knee'' frequency, or frequency where the PSD is a minimum.  The quantities $I(7)$ and $\Delta$ are determined from the Fisher matrix inversion and are largely independent of either $S_0$ or $f_0$, or of the SNR of the signal.  In any case, the bound is only weakly dependent on these variables.  The ratio $D/(1+Z)D_L$ is weakly dependent on distance, reflecting the fact that the effect of the massive graviton and the estimation errors both grow with distance.   As it turns out, the factor $S_0^{1/4} f_0^{1/3}$ is roughly the same for LISA as it is for AdvLIGO, and thus the bound on $\lambda_g$ is almost entirely proportional to the chirp mass of the source.

Though the bound decreases with increase in the mass ratio,
the improvement due to the inclusion of higher harmonics is more profound with increasing mass ratio. The
amplitude corrections to the higher harmonics are proportional to the mass
asymmetry of the binary for all odd PN amplitude orders (see e.g. Eqs~(8.9)-(8.10) of \cite{BFIS08}) which include the leading correction at 0.5PN order, and hence the above observation is not surprising.  It should also be borne in mind that, for very large mass ratios, the systems are more like Extreme Mass Ratio Inspirals, and many other PN effects will
have to be accounted for in order to detect them. 
As in the case of the
ground-based detectors, use of the FWF increases the range of masses over which stringent
bounds can be put on the $\lambda_g$ term.  This range of masses ($10^6M_\odot$- a few times $10^7M_\odot$) is very much in the astrophysically interesting regime.
 
As mentioned earlier, we have not taken into account the orbital motion of
LISA in deriving these estimates. However, the addition of angular parameters
should not strongly affect the reported bounds because they are rather uncorrelated 
with the $\beta_g$ parameter.  On the contrary, inclusion of LISA's orbital motion could enhance
the SNRs and thereby improve the bounds.

\section{Conclusions}
\label{sec:conclusions}

We have shown that the use of amplitude corrected PN waveforms which incorporate higher harmonics
of the orbital phase in constructing templates for matched filtering (especially
for parameter estimation) can provide more stringent bounds on the
massive graviton parameter.  Third generation ground-based interferometers such an Einstein telescope, and the space-based LISA would almost continuously cover the BH inspirals ranging from stellar mass binaries of a few tens of solar mass up to 
supermassive BHs of $10^8 M_\odot$. 

Table 1 summarizes the typical bounds on the graviton Compton wavelength achievable from different interferometric gravitational-wave detectors incorporating higher harmonics  into the gravitational waveform templates. Also shown are
typical orders of magnitudes of the observational mass range and the signal to noise ratio (SNR). The mass range varies with the chosen low frequency cut-off of the detector.  For ET, the mass range is between $1 M_\odot$ and $10^{3}M_\odot$ or $10^{4}M_\odot$, depending on whether the seismic cut-off chosen is 10 Hz or 1 Hz, respectively.  For LISA, the
mass range is between $10^4 M_\odot$ and $10^{8}M_\odot$ or $10^{9}M_\odot$, depending on whether the low frequency cut-off of $10^{-4}$ or $10^{-5}$ Hz, respectively .   Numbers for the mass range and SNR have been taken from the literature~\cite{ChrisAnand06,AISS07}.  The best bounds would
be from the proposed space-based interferometer, LISA.

\begin{center}
\begin{table}[t]
\begin{tabular}{ |l|| l l l| }
\hline
 Detector & Mass Range ($M_\odot$) & SNR & $\lambda_g$ (km) \\
\hline
  AdvLIGO & 1--400 & few tens & $10^{12}$\\
  Einstein Telescope & 1--$10^{3-4}$ & few tens--few hundreds &$10^{13-14}$\\
  LISA & $10^4$--$10^{8-9}$ & few thousands & $10^{16}$\\
\hline
\end{tabular}
\caption{Summary of bounds on the graviton Compton wavelength from various gravitational wave detectors using full waveform templates}
\end{table}
\end{center}

\ack
We thank Adamantios Stavridis for many useful discussions. We thank P. Ajith
for discussions during which we spotted a typo in our code.
This work was supported in part by the National Science Foundation, Grant No.\ PHY 06--52448, the National Aeronautics and Space Administration, Grant No.\
NNG-06GI60G, and the Centre National de la Recherche Scientifique, Programme Internationale de la Coop\'eration Scientifique (CNRS-PICS), Grant No. 4396.  KGA is a VESF fellow at the European Gravitational Observatory.

\section{References}
\bibliographystyle{iopart-num.bst}
\bibliography{/home/arun/tphome/arun/ref-list}

\providecommand{\newblock}{}
\begin{thebibliography}{10}
\expandafter\ifx\csname url\endcsname\relax
  \def\url#1{{\tt #1}}\fi
\expandafter\ifx\csname urlprefix\endcsname\relax\def\urlprefix{URL }\fi
\providecommand{\eprint}[2][]{\url{#2}}

\bibitem{SathyaSchutzLivRev09}
{Sathyaprakash, B S and Schutz, B F} 2009 {\em Living Rev. Rel.\/} {\bf 12} 2
  (\textit{Preprint} \eprint{arxiv:0903.0338[gr-qc]}),
URL:http://www.livingreviews.org/lrr-2009-2 (cited on 9 June 2009)

\bibitem{ligo}
\url{http://www.ligo.caltech.edu}

\bibitem{virgo}
\url{http://www.virgo.infn.it}

\bibitem{ET}
\url{http://www.et-gw.eu/}

\bibitem{Danzmann97}
Danzmann K 1997 {\em Class.~Quantum~ Grav.\/} {\bf 14} 1399

\bibitem{Will98}
Will C~M 1998 {\em Phys.~ Rev.~D\/} {\bf 57} 2061 (\textit{Preprint}
  \eprint{gr-qc/9709011})

\bibitem{WillYunes04}
Will C~M and Yunes N 2004 {\em Class. Quantum Grav.\/} {\bf 21} 4367
  (\textit{Preprint} \eprint{gr-qc/0403100})

\bibitem{BBW05a}
{Berti} E, {Buonanno} A and {Will} C~M 2005 {\em Phys.~Rev.~D\/} {\bf 71}
  084025--+ (\textit{Preprint} \eprint{gr-qc/0411129})

\bibitem{BBW05b}
Berti E, Buonanno A and Will C~M 2005 {\em Class. Quantum Grav.\/} {\bf 22}
  S943 (\textit{Preprint} \eprint{gr-qc/0504017})

\bibitem{Zakharov70}
Zakharov V~I 1970 {\em J.~Exp.~Theor.~Phys.~Lett.\/} {\bf 12} 312

\bibitem{Veltman70}
Van~Dam H and Veltman M~J~G 1970 {\em Nucl.~Phys.~B\/} {\bf 22} 397

\bibitem{DIS98}
Damour T, Iyer B~R and Sathyaprakash B~S 1998 {\em Phys. Rev. D\/} {\bf 57}
  885--907

\bibitem{Bliving}
Blanchet L 2006 {\em Living Rev. Rel.\/} {\bf 9} 4 (\textit{Preprint}
  \eprint{gr-qc/0202016})\\
URL: http://www.livingreviews.org/lrr-2006-4 (cited on 9 June, 2009)

\bibitem{SinVecc00a}
Sintes A~M and Vecchio A 2000 {\em Rencontres de Moriond:Gravitational waves
  and experimental gravity\/} ed Dumarchez J (Frontières, Paris)
  (\textit{Preprint} \eprint{gr-qc/0005058})

\bibitem{SinVecc00b}
{Sintes} A~M and {Vecchio} A 2000 {\em Third Amaldi conference on Gravitational
  Waves\/} ed {Meshkov} S (American Institute of Physics Conference Series) p
  403 (\textit{Preprint} \eprint{gr-qc/0005059})

\bibitem{MH02}
Moore T~A and Hellings R~W 2002 {\em Phys. Rev.~D\/} {\bf 65} 062001

\bibitem{HM03}
Hellings R~W and Moore T~A 2003 {\em Class.~Quant.~Grav.\/} {\bf 20} S181
  (\textit{Preprint} \eprint{gr-qc/0207102})

\bibitem{Chris06}
Van Den~Broeck C 2006 {\em Class.~Quantum Grav.\/} {\bf 23} L51
  (\textit{Preprint} \eprint{gr-qc/0604032})

\bibitem{ChrisAnand06}
Van Den~Broeck C and Sengupta A 2007 {\em Class.~Quantum Grav.\/} {\bf 24} 155
  (\textit{Preprint} \eprint{gr-qc/0607092})

\bibitem{ChrisAnand06b}
{Van Den Broeck} C and {Sengupta} A~S 2007 {\em Class.~Quantum Grav.\/} {\bf
  24} 1089--1113 (\textit{Preprint} \eprint{gr-qc/0610126})

\bibitem{AISS07}
{Arun} K~G, {Iyer} B~R, {Sathyaprakash} B~S and {Sinha} S 2007 {\em
  Phys.~Rev.~D\/} {\bf 75} 124002 (\textit{Preprint} \eprint{arXiv:0704.1086
  [gr-qc]})

\bibitem{AISSV07}
Arun K~G, Iyer B~R, Sathyaprakash B~S, Sinha S and Broeck C~V~D 2007 {\em
  Phys.~Rev.~D\/} {\bf 76} 104016 (\textit{Preprint} \eprint{arXiv:0707.3920
  [astro-ph]})

\bibitem{TriasSintes07}
Trias M and Sintes A~M 2008 {\em Phys. Rev. D\/} {\bf 77} 024030
  (\textit{Preprint} \eprint{arXiv:0707.4434 [gr-qc]})

\bibitem{PorterCornish08}
Porter E~K and Cornish N~J 2008 {\em Phys. Rev. D\/} {\bf 78} 064005
  (\textit{Preprint} \eprint{arxiv:0804.0332[gr-qc]})

\bibitem{AMVISS08}
Arun K~G, Mishra C, Van Den~Broeck C, Iyer B~R, Sathyaprakash B~S and Sinha S
  2009 {\em Class. Quant. Grav. {\bf 26} 094021}  (\textit{Preprint} \eprint{arXiv:0810.5727})

\bibitem{BIWW96}
Blanchet L, Iyer B~R, Will C~M and Wiseman A~G 1996 {\em Class. Quantum
  Grav.\/} {\bf 13} 575--584 (\textit{Preprint} \eprint{gr-qc/9602024})

\bibitem{ABIQ04}
Arun K~G, Blanchet L, Iyer B~R and Qusailah M~S~S 2004 {\em Class. Quantum
  Grav.\/} {\bf 21} 3771 erratum-ibid. {\bf 22}, 3115 (2005) (\textit{Preprint}
  \eprint{gr-qc/0404185})

\bibitem{KBI07}
Kidder L~E, Blanchet L and Iyer B~R 2007 {\em Class. Quant. Grav.\/} {\bf 24}
  5307 (\textit{Preprint} \eprint{arXiv:0706.0726 [gr-qc]})

\bibitem{BFIS08}
Blanchet L, Faye G, Iyer B~R and Sinha S 2008 {\em Class. Quantum. Grav.\/}
  {\bf 25} 165003 (\textit{Preprint} \eprint{arXiv:0802.1249})

\bibitem{BDIWW95}
Blanchet L, Damour T, Iyer B~R, Will C~M and Wiseman A~G 1995 {\em Phys. Rev.
  Lett.\/} {\bf 74} 3515--3518 (\textit{Preprint} \eprint{gr-qc/9501027})

\bibitem{B96}
Blanchet L 1996 {\em Phys. Rev. D\/} {\bf 54} 1417--1438 {Erratum-ibid.{\bf
  71}, 129904(E) (2005)} (\textit{Preprint} \eprint{gr-qc/9603048})

\bibitem{BFIJ02}
Blanchet L, Faye G, Iyer B~R and Joguet B 2002 {\em Phys. Rev. D\/} {\bf 65}
  061501(R) {Erratum-ibid~{\bf 71}, 129902(E) (2005)} (\textit{Preprint}
  \eprint{gr-qc/0105099})

\bibitem{BDEI04}
Blanchet L, Damour T, Esposito-Far{\`e}se G and Iyer B~R 2004 {\em Phys. Rev.
  Lett.\/} {\bf 93} 091101 (\textit{Preprint} \eprint{gr-qc/0406012})

\bibitem{AISS05}
Arun K~G, Iyer B~R, Sathyaprakash B~S and Sundararajan P~A 2005 {\em
  Phys.~Rev.~D\/} {\bf 71} 084008 erratum-ibid. ~{\bf D } 72, 069903 (2005)
  (\textit{Preprint} \eprint{gr-qc/0411146})

\bibitem{ETPSD}
https://workarea.et-gw.eu/et/WG4-Astrophysics/base-sensitivity/wg4-assumptions%
.pdf

\bibitem{ABFO08}
Arun K~G, Buonanno A, Faye G and Ochsner E 2009, {\em Phys. Rev. D {\bf 79} 104023}  (\textit{Preprint}
  \eprint{arxiv:0810.5336[gr-qc]})

\bibitem{DPK99}
Droz S, Knapp D~J, Poisson E and Owen B~J 1999 {\em Phys. Rev. D\/} {\bf 59}
  124016

\bibitem{DIS00}
Damour T, Iyer B~R and Sathyaprakash B~S 2000 {\em Phys. Rev. D\/} {\bf 62}
  084036 (\textit{Preprint} \eprint{gr-qc/0001023})

\bibitem{CF94}
Cutler C and Flanagan E 1994 {\em Phys. Rev. D\/} {\bf 49} 2658--2697

\bibitem{PW95}
Poisson E and Will C 1995 {\em Phys. Rev. D\/} {\bf 52} 848--855

\bibitem{Vallisneri07}
Vallisneri M 2008 {\em Phys. Rev. D\/} {\bf 77} 042001 (\textit{Preprint}
  \eprint{gr-qc/0703086})

\end{thebibliography}
\end{document}